\documentclass[aps,a4paper,superscriptaddress,twocolumn]{revtex4}
\usepackage{tensor}
\usepackage{graphicx}
\usepackage{amsmath}
\usepackage{amssymb}
\usepackage{enumerate}
\usepackage{subfigure}
\usepackage{tabularx}
\usepackage[colorlinks=true, pdfstartview=FitV, linkcolor=blue, citecolor=red, urlcolor=black, breaklinks=true]{hyperref}
\newcommand{\be}{\begin{equation}}
\newcommand{\ee}{\end{equation}}
\newcommand{\ben}{\begin{eqnarray}}
\newcommand{\een}{\end{eqnarray}}
\newcommand{\bes}{\begin{subequations}}
\newcommand{\ees}{\end{subequations}}
\def\bal#1\eal{\begin{align}#1\end{align}}

\newcommand{\sech}{{\rm sech}}
\newcommand{\LL}{{\mathcal L}}

\newcommand{\veps}{\varepsilon}
\begin{document}
\title{Magnetic monopoles with internal structure}
\author{D. Bazeia}\affiliation{Departamento de F\'\i sica, Universidade Federal da Para\'\i ba, 58051-970 Jo\~ao Pessoa, PB, Brazil}
\author{M.A. Marques}\affiliation{Departamento de F\'\i sica, Universidade Federal da Para\'\i ba, 58051-970 Jo\~ao Pessoa, PB, Brazil}
\author{R. Menezes}\affiliation{Departamento de Ci\^encias Exatas, Universidade Federal da Para\'{\i}ba, 58297-000 Rio Tinto, PB, Brazil}\affiliation{Departamento de F\'\i sica, Universidade Federal da Para\'\i ba, 58051-970 Jo\~ao Pessoa, PB, Brazil}
\begin{abstract}
We investigate the presence of magnetic monopoles in a model that extends the non Abelian model originally studied by 't Hooft and Polyakov with the inclusion of an extra neutral field. The investigation includes modifications of the dynamics of the gauged fields, and the main results unveil a route to construct solutions that engender internal structure and live in a compact space. 
\end{abstract}
\date{\today}
\maketitle

\section{Introduction}

 Topological structures appear in a diversity of contexts in high energy physics \cite{V,MS,Va,S,W}. Among the several possible configurations, perhaps the most known structures are kinks in the line, vortices in the plane and magnetic monopoles in the space. In general, kinks are described by real scalar fields, vortices by charged scalar coupled to an Abelian gauge field, and monopoles require non Abelian gauge fields. 

In this work we concentrate on magnetic monopoles and deal with a system similar to the model investigated before in Refs.~\cite{tH,P}. As one knows, magnetic monopoles are of current interest in physics \cite{V,MS,S} and some studies on them have been carried out recently, for instance, in \cite{Sh,Va1,Ya,H,C} and in references therein. In the current work, we enlarge the local $SU(2)$ symmetry described in \cite{tH,P} to accommodate an extra neutral field $\phi$ which engenders the global $Z_2$ symmetry. In this sense, we study a model that engenders the $SU(2)\times Z_2$ symmetry, with the main motivation coming from the recent work \cite{bmm}, in which we studied vortices in a model with the local $U(1)$ symmetry enlarged to describe the $U(1)\times Z_2$ symmetry, with the additional $Z_2$ symmetry describing the inclusion of a neutral scalar field. 

The study of monopoles is much more intricate since it usually requires the presence of non Abelian gauge symmetry and several degrees of freedom. For this reason, in the current work we study a simple extension of the model considered in \cite{tH,P}, with the inclusion of the neutral field and some modification in the Lagrangian of the original model, as we explain below. In the recent work \cite{kvm} we have investigated issues related to the construction of twinlike models in the presence of kinks, vortices and monopoles, in the line, plane and space, respectively, and the study of monopoles motivated us to elaborate the current study. The investigation is also motivated by the work of Ref.~\cite{shi}, which shows how a conventional topological defect can acquire non-Abelian moduli localized on its world sheet, and by the recent work \cite{mono}, in which the authors investigate the presence of color-magnetic defects in dense quark matter, as in the interior of compact stars, which may contribute to  produce detectable gravitational waves. In condensed matter, another motivation to study monopoles appears with the emergency of magnetic monopoles \cite{mmsi} in a class of exotic magnets known collectively as spin ice \cite{si0,si1,si2}, in which the dipole moment of the underlying structure may fractionalize into monopoles. As shown in \cite{mmed}, for instance, magnetic monopoles in spin ice, besides having magnetic charge, may also have an electric dipole. We then say that magnetic monopoles may engender internal structure.

In order to ease the current investigation, we require that the equations of motion are solved by first order differential equations, which appear from the Bogomol'nyi-Prasad-Sommerfield (BPS) procedure \cite{PS,B}. This lead us with stable finite energy configurations, and we describe the possibility to add internal structure and make them compact. Below we introduce the general model, explain the procedure, investigate and solve some specific models with focus on the presence of internal structure, and then close the work adding some new possibilities of study.

\section{The General model}
Let us consider an extension of the model considered in \cite{tH,P}. To do this, we work in $(3,1)$ spacetime dimensions and take the Lagrangian density
\be\label{lmodel}
\begin{aligned}
\LL &= - \frac{1}{4}P(\phi)F^{a}_{\mu\nu}F^{a\mu\nu} -\frac{M(\phi)}{2} D_\mu \chi^a D^\mu \chi^a \\
&\hspace{4mm} - \frac12\partial_\mu\phi\partial^\mu\phi- V(\phi,|\chi|).
\end{aligned}
\ee
Here $\phi$ is a neutral scalar field, and $\chi^a$ is a triplet of scalar fields in the adjoint representation that couples to the gauge field $A^a_\mu$ under the $SU(2)$ symmetry. The general model is an extension of a model which is known to admit solutions of the monopole type. We also have $|\chi|=\sqrt{\chi^a\chi^a}$, $D_\mu \chi^a = \partial_\mu\chi^a + g\veps^{abc}A^b_\mu\chi^c$ and $F^a_{\mu\nu} = \partial_\mu A^a_\nu - \partial_\nu A^a_\mu + g\veps^{abc}A^b_\mu A^c_\nu$. The function $P(\phi)$ is generalized magnetic permeability and $M(\phi)$ modifies the dynamics of the field $\chi^a$; they only depend on the neutral field $\phi$, and we suppose that they are non negative functions.  Moreover, $g$ is the coupling constant between the $A_\mu$ and $\chi$ fields, the indices $a,b,c=1,2,3$ stand for the $SU(2)$ symmetry of the fields and the greek letters $\mu,\nu=0,1,2,3$ represent the spacetime indices. 

The metric tensor is $\eta_{\mu\nu} = \textrm{diag}(-,+,+,+)$ and we use natural units, that is, $\hbar=c=1$. Previous investigations on generalized models searching for magnetic monopoles in the presence of magnetic permeability and modification of the dynamics of the scalar field have been considered before in \cite{dH,BL}, for instance; here, however, we approach the problem from a well distinct perspective.

The equations of motion associated to the Lagrangian density \eqref{lmodel} are
\bes\label{geom}
\begin{align}
\partial_\mu\partial^\mu\phi &= \frac14 P_\phi F^a_{\mu\nu}F^{a\mu\nu} + V_\phi,\\
 D_\mu\left(M(\phi) D^\mu \chi^a\right) &= \frac{\chi^a}{|\chi|} V_{|\chi|},\\ \label{meqsc}
 D_\mu\left(P(\phi) F^{a\mu\nu}\right) &=  gM(\phi)\veps^{abc}\chi^b D^\nu \chi^c,
\end{align}
\ees
where $D_\mu F^{a\mu\nu} = \partial_\mu F^{a\mu\nu} + g\veps^{abc}A^b_\mu F^{c\mu\nu}$, $V_\phi = d V/d\phi$,
$V_{|\chi|} = d V/d|\chi|$, and $P_\phi=dP/d\phi$.
To search for the monopole, we consider static configurations and take $A_0=0$. Also, we consider the relevant fields in the form
\be\label{ansatz}
\phi=\phi(r),\quad\chi^a = \frac{x_a}{r} H(r), \quad A_i^a = \veps_{aib}\frac{x_b}{gr^2}(1-K(r)),\;
\ee
with the boundary conditions
\be\label{bcond}
\begin{aligned}
\phi(0)&=\phi_0, & H(0)&=0, & K(0)&=1,\; \\
\phi(\infty)&=\phi_\infty, & \lim_{r\to\infty}{H(r)} &= \pm\eta, & \lim_{r\to\infty}{K(r)} &= 0.\;
\end{aligned}
\ee
The values $\phi_0$ and $\phi_\infty$ are constants related to the behavior of the neutral field, and $\eta$ is a constant that control the asymptotic value of $H(r)$. In this case, the equations of motion \eqref{geom} become
\bes\label{geomansatz}
\begin{align}
\phi^{\prime\prime} &= \frac{P_\phi}{2} \left(\frac{2{K^\prime}^2}{g^2r^2}+\frac{(1-K^2)^2}{g^2r^4}\right) + V_\phi\\
 \left(r^2M H^\prime\right)^\prime &= 2M H K^2 + r^2V_{H},   \\
r^2\left(P K^\prime\right)^\prime &= K\left(M g^2r^2H^2 -P(1-K^2)\right),
\end{align}
\ees
where the prime denotes derivative with respect to $r$. The above equations are coupled second order differential equations. In order to get first order equations, we follow the BPS procedure to minimize the energy. For the fields \eqref{ansatz}, the energy density is conserved and can be calculated standardly; it is given by
\be\begin{aligned}\label{rho}
\rho &= \frac{P(\phi)}{2} \left(\frac{2{K^\prime}^2}{g^2r^2} + \frac{(1-K^2)^2}{g^2r^4}\right) \\
     &\hspace{4mm}+ \frac{M(\phi)}{2}\left({H^\prime}^2 + \frac{2H^2K^2}{r^2}\right) + \frac12 {\phi^\prime}^2 + V(\phi,|\chi|).
\end{aligned}
\ee
In the case $M(\phi)=1/P(\phi)$, we can introduce an auxiliar function $W(\phi)$ and write the above equation as
\begin{align}\label{bps1}
\rho &= \frac12\left(\phi^\prime \mp \frac{W_\phi}{r^2}\right)^2 +  P(\phi)\left( \frac{K^\prime}{gr} \pm \frac{HK}{rP(\phi)}\right)^2\nonumber\\
     &\hspace{4mm} \!+\!\frac{1}{2P(\phi)} \left({H^\prime} \mp \frac{P(\phi)(1\!-\!K^2)}{gr^2}\right)^2\!\! +\! V(\phi,|\chi|)-\frac12\!\frac{W^2_\phi}{r^4} \nonumber\\
     &\hspace{4mm} + \frac{1}{r^2}\left(W + \frac{\left(1-K^2\right)H}{g}\right)^\prime.
\end{align}
If we further suppose that the potential depends only on $\phi$, in the form
\be\label{potm}
V(\phi) = \frac12\frac{W^2_\phi}{r^4},
\ee
we get the interesting result
\be
\begin{aligned}
\rho &= \frac12\left(\phi^\prime \mp \frac{W_\phi}{r^2}\right)^2 + P(\phi)\!\left(\frac{K^\prime}{gr} \pm \frac{HK}{rP(\phi)}\right)^2\\
     &\hspace{4mm} + \frac{1}{2P(\phi)} \left({H^\prime} \mp \frac{P(\phi)(1-K^2)}{gr^2}\right)^2 \\
     &\hspace{4mm} + \frac{1}{r^2}\!\left(W + \frac{\left(1-K^2\right)H\!}{g}\right)^\prime.
\end{aligned}
\ee
We then note that the above three first terms are non-negative and therefore, the energy is bounded, i.e., $E\geq E_B$, where
\be\label{ebogo}
E_B = 4\pi\left|W(\phi_\infty) - W(\phi_0)\right| + \frac{4\pi\eta}{g}.
\ee

The solutions that minimize the energy of the system are obtained when they satisfy the first order equations
\be\label{fophi}
\phi^\prime =\pm \frac{W_\phi}{r^2},
\ee
and
\bes\label{fom}
\bal
H^\prime &=\pm \frac{P(\phi)(1-K^2)}{gr^2},\\
K^\prime &=\mp \frac{gHK}{P(\phi)}.
\eal
\ees
In this case, the energy is minimized to $E=E_B$, given by Eq.~\eqref{ebogo}. One can show that the first order equations \eqref{fophi} and \eqref{fom} are compatible with the equations of motion \eqref{geomansatz}. We also notice that the two set of first order equations \eqref{fom} are linked by the change $H(r)\to -H(r)$.

We notice that the above BPS procedure which gives the equations $\eqref{bps1}$-$\eqref{fom}$ naturally adds the radial factor in the potential \eqref{potm} and in the first order equations \eqref{fophi}, and this reminds us very much of the procedure used in Ref.~\cite{bmmprl} to circumvent the Derrick-Hobart theorem \cite{D,Ho}. This is an interesting issue, and it also appeared before in \cite{bmm}, in the study of vortex configurations in the plane. In the case of monopoles, we also notice that the potential in Eq.~\eqref{potm} does not depend on the triplet scalar $\chi$, in accordance with the result of Ref.~\cite{PS}. 

The equations for the neutral scalar field are decoupled from the other two equations \eqref{fom}, which may describe the magnetic monopole. This lead us to think of the neutral field as a source field to generate the monopole, so we refer to $\phi$ as the source field. In this sense, in order to find monopole solutions, one first solve \eqref{fophi} and then introduce a generalized magnetic permeability $P(\phi)$ to deal with \eqref{fom}. Before doing this, however, we note that the above first order equations allow us to write the energy density in Eq.~\eqref{rho} as $\rho = \rho_m + \rho_s$, with $\rho_m$ representing the contribution of the monopole and $\rho_s$ identifying the contribution of the source field to the energy density. We have
\bes
\bal\label{rhom}
\rho_m &=  \frac{2P(\phi){K^\prime}^2}{g^2r^2} + \frac{{H^\prime}^2}{P(\phi)}, \\ \label{rhos}
\rho_s &= {\phi^\prime}^2,
\eal
\ees
showing that the energy densities of the neutral field does not depend on the other fields. As shown by the BPS procedure, we see from Eq.~\eqref{ebogo} that the total energy changes only with $W(\phi)$. The monopole energy is fixed, $E_m=4\pi\eta/g$, regardless the magnetic permeability. On the other hand, the source field energy changes with the function $W(\phi)$ and is given by $E_s=4\pi|W(\phi_\infty) - W(\phi_0)|$.

From now on, we work with dimensionless fields, keeping in mind that the rescale
\be
\begin{aligned}
	\chi^a &\to \eta \chi^a, & \phi &\to \eta\phi, & A^a_\mu &\to \eta A^a_\mu, \\
	 r &\to (g\eta)^{-1} r, & W_\phi &\to g^{-1} W_\phi, & \LL &\to g^2\eta^4 \LL
\end{aligned}
\ee 
can be done. We also take $g,\eta=1$, for simplicity, and illustrate the general procedure investigating some specific models below. 

\section{Specific examples} 

We have distinct possibilities to deal with the source field, and here we consider some specific cases. 

\subsection{A simple model}

\begin{figure}[t!]
\centering
\includegraphics[width=4.2cm]{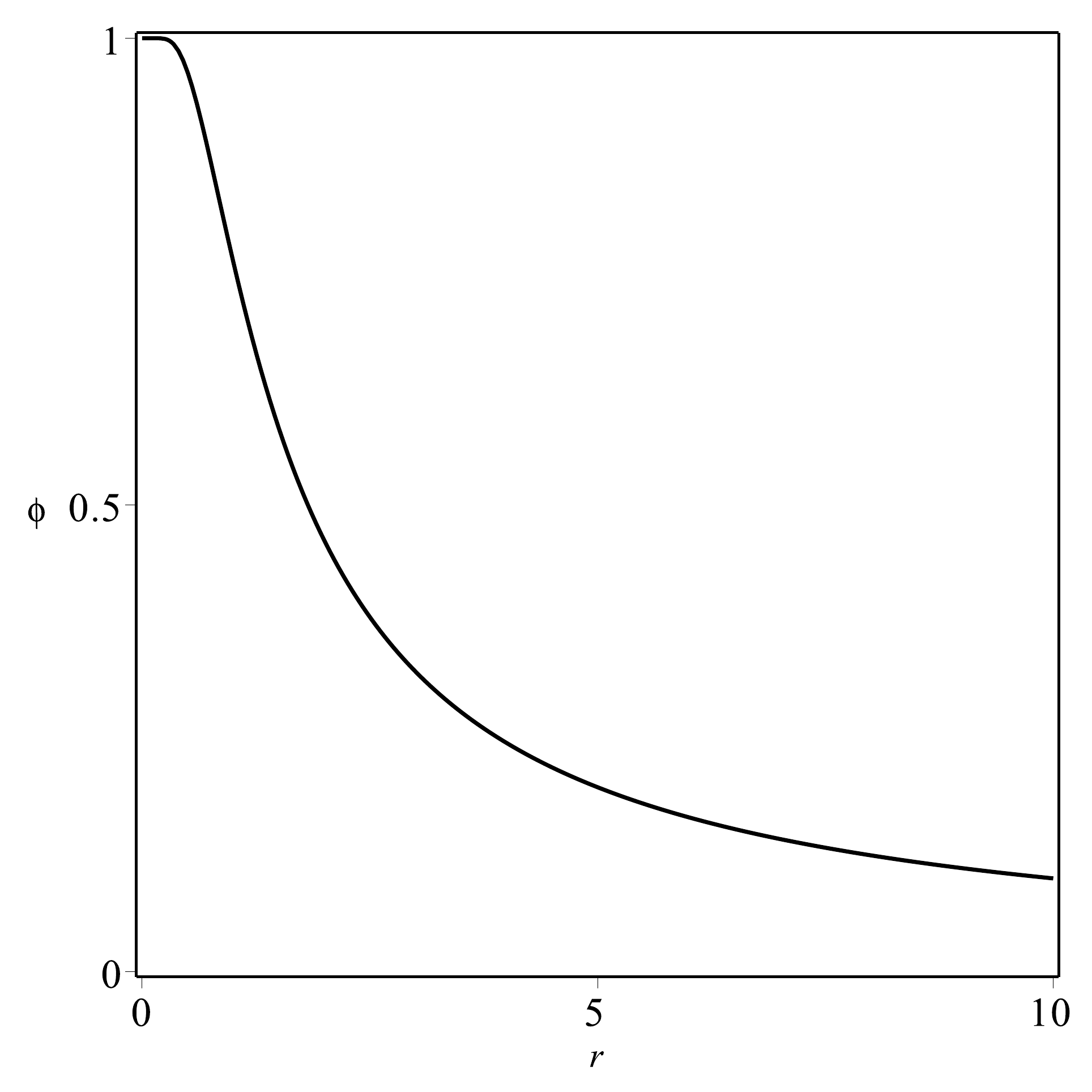}
\includegraphics[width=4.2cm]{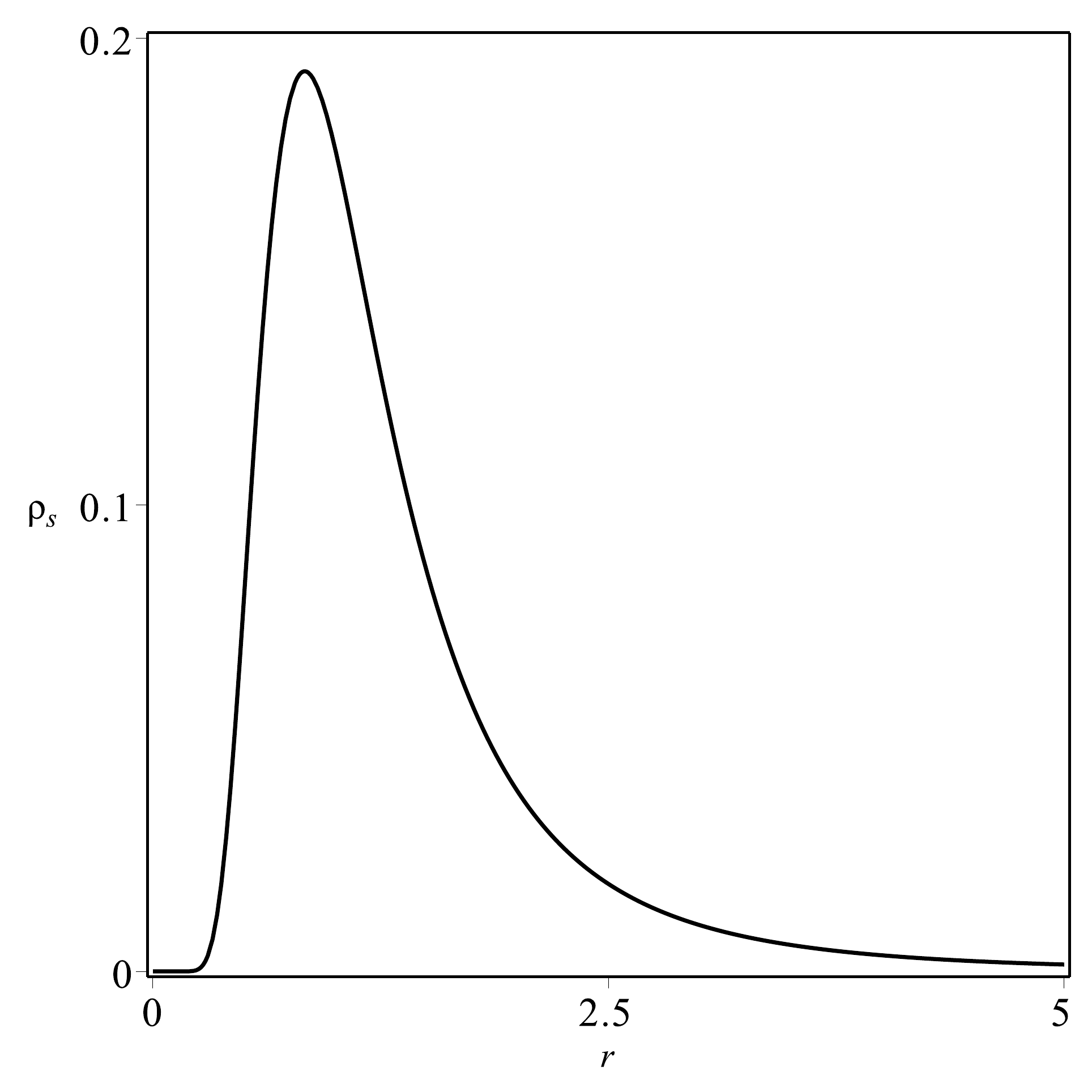}
\caption{The solution in Eq.~\eqref{solphi4} (left) and its energy density \eqref{rhophi4} (right)}
\label{fig1}
\end{figure}

Let us first consider the case in which the source field is described by the function 
\be
W(\phi) = \frac13\phi^3 - \phi.
\ee
The associated potential has minima at $\phi=\pm1$ and a maximum at $\phi=0$. The first order equation \eqref{fophi} with the above choice admits the intersting analytic solution
\be\label{solphi4}
\phi(r) = \tanh\left(\frac1r\right).
\ee
It connects the minimum $\phi=1$ at the origin to the local maximum $\phi=0$ asymptotically, which we use as $\phi_0$ and $\phi_{\infty}$; see the boundary conditions for the neutral field in \eqref{bcond}. In this case, the energy density \eqref{rhos} becomes
\be\label{rhophi4}
\rho_s(r) = \frac{1}{r^4}\,\sech^4\left(\frac1r\right).
\ee
A direct integration shows that the energy is $E_s=8\pi/3$, which is the result expected from Eq.~\eqref{ebogo}. In Fig.~\ref{fig1}, we depict the solution \eqref{solphi4} and the energy density \eqref{rhophi4}.

\begin{figure}[htb!]
\centering
\includegraphics[width=4.2cm]{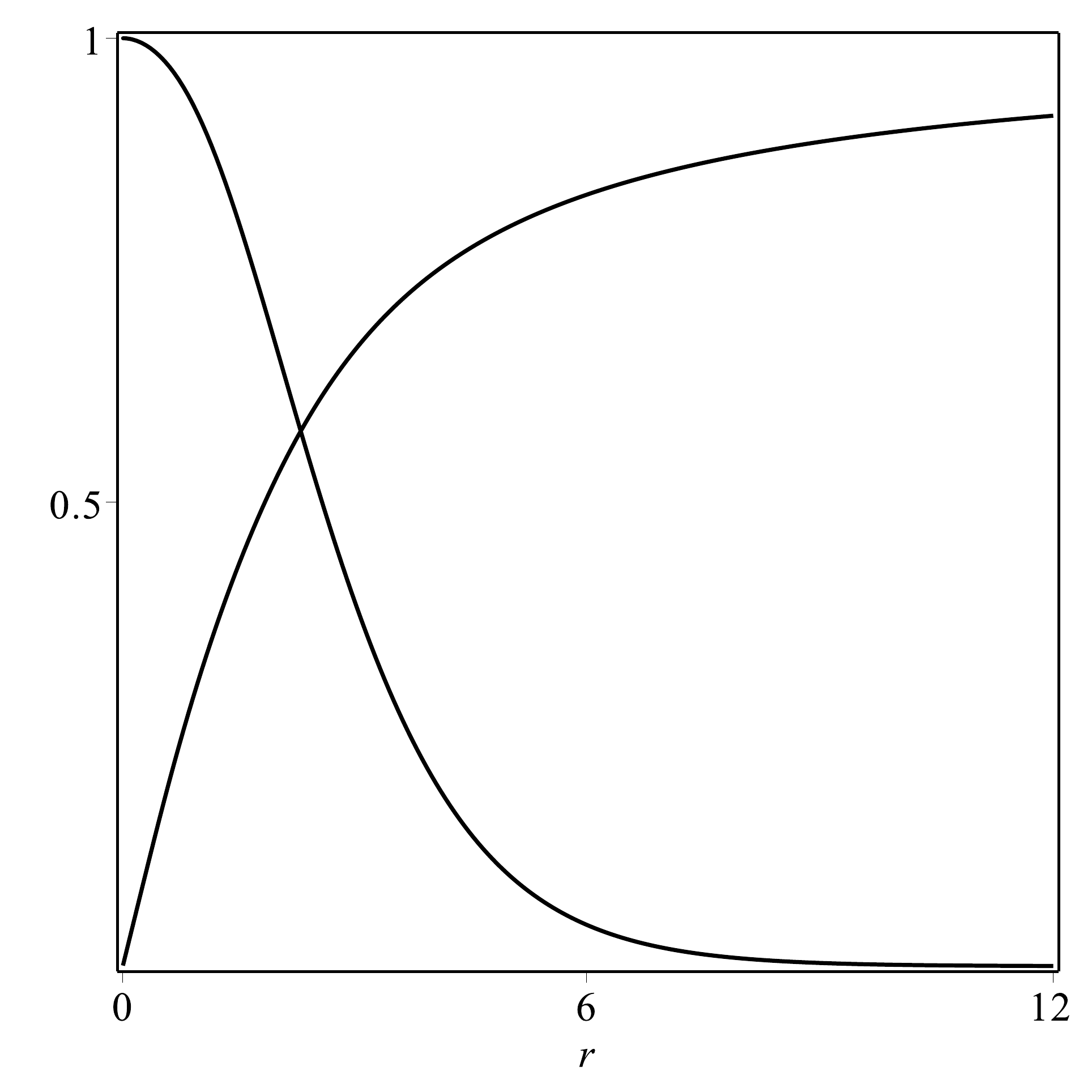}
\includegraphics[width=4.2cm]{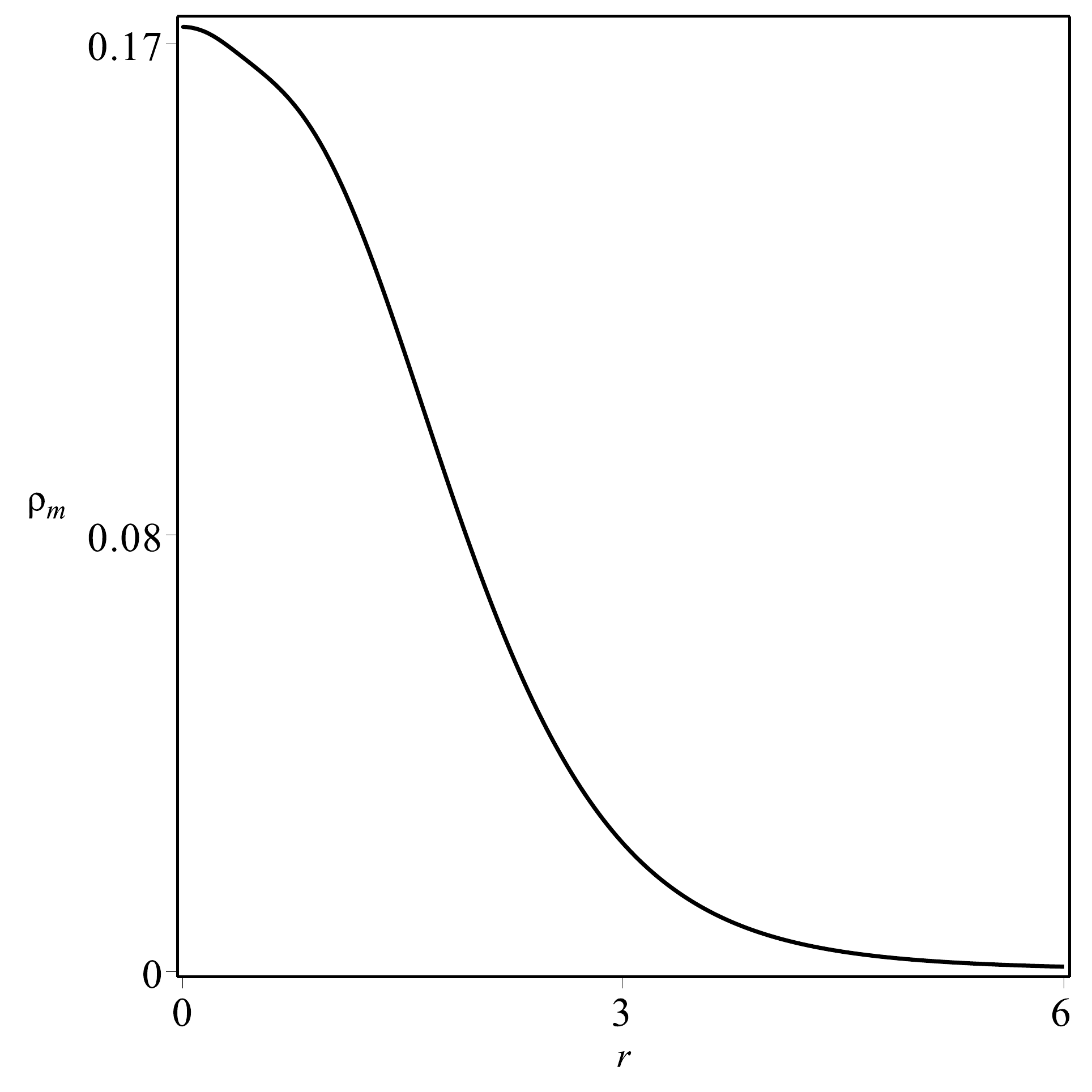}
\caption{The solutions $H(r)$ (ascending line) and $K(r)$ (descending line) of Eqs.~\eqref{fom1} (left) with the source field given by \eqref{solphi4} and the corresponding energy density of the monopole (right).}
\label{fig2}
\end{figure}

\begin{figure}[htb!]
\centering
\includegraphics[width=4.8cm]{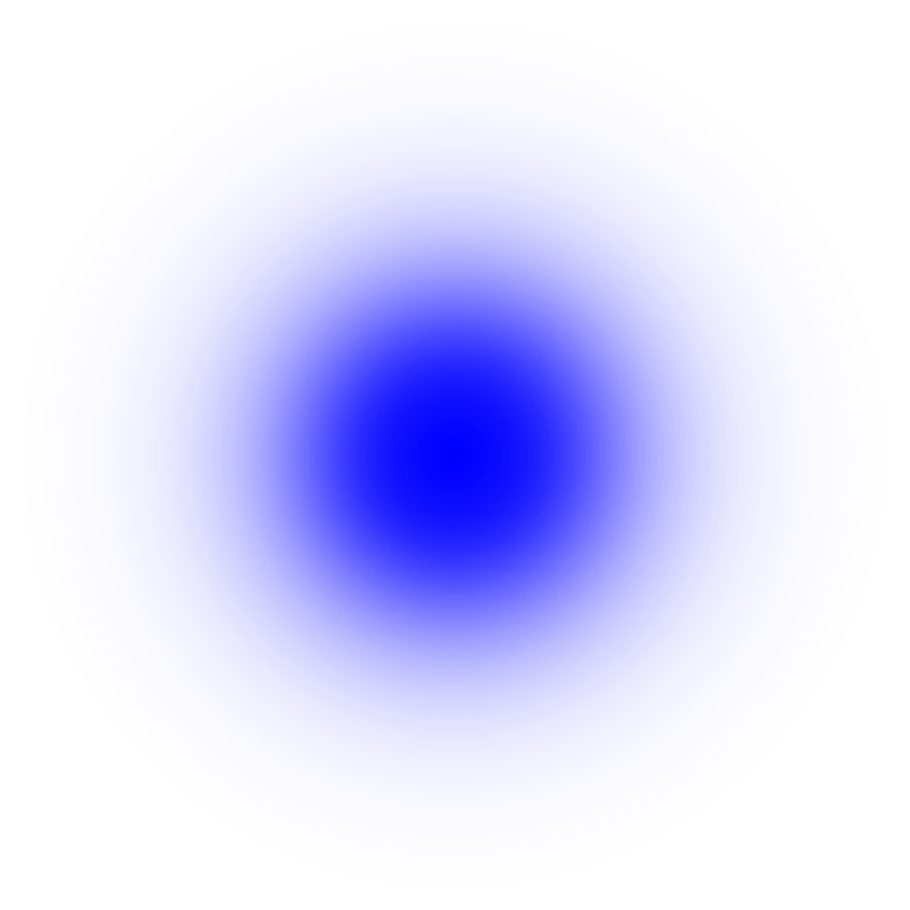}
\caption{The monopole energy density of the solution of Eqs.~\eqref{fom1} with the source field given by Eq.~\eqref{solphi4}. We depict a planar section of the energy density passing through the center of the structure, with the darkness of the color directly related to the intensity of the energy density.}
\label{fig3}
\end{figure}

We use this source to generate the magnetic permeability controlled by the function 
\be\label{perm}
P(\phi)=1+\phi^2.
\ee
In this case, the first order equations \eqref{fom} become
\bes\label{fom1}
\bal
H^\prime &= \frac{(1+\phi^2)(1-K^2)}{r^2},\\
K^\prime &= -\frac{HK}{(1+\phi^2)},
\eal
\ees
with $\phi$ given by Eq.~\eqref{solphi4}. These equations give rise to the magnetic monopole. In Fig.~\ref{fig2} we display the numerical solutions and the energy density of the monopole, which can also be seen in Fig.~\ref{fig3} as a planar section passing through the center of the structure. A numerical integration of the energy density leads to energy $E_m=4\pi$, in accordance with Eq.~\eqref{ebogo} for $\eta,g=1$.

As we see from Figs.~\ref{fig2} and \ref{fig3}, the energy density of the magnetic monopole has standard profile, diminishing from a given value at the origin toward zero asymptotically. We have considered another magnetic permeability, using $P(\phi)=\phi^2$ to see if it could change the above features, but the monopole solution presented similar qualitative profile. 

\subsection{Other models} 

To change the profile of the magnetic monopole, we have to investigate other models, and here we further illustrate the construction of solutions considering the source field to be driven by the function
\be\label{anomod}
W(\phi) = \frac37\phi^{7/3} - \frac35\phi^{5/3}.
\ee
This is a particular case of the model considered before in Eq. (6) of Ref.~\cite{bmmprl} for the value $p=3$, and it allows that we explore compact solutions. Thus, before going on and search for explicit solutions, let us further comment on the presence of compact solutions. In the recent years, we have been investigating compact structures in a diversity of contexts, including compact kinks \cite{CK}, lumps \cite{CL}, Q-balls \cite{CQ} and vortices \cite{CV1,CV2}, and these results motivate us to investigate the possibility of finding magnetic monopoles of the compact type as well, which we describe below.

We now return to the $W$ introduced in Eq.~\eqref{anomod} to see that the minima of $W_\phi^2$ are located at $\phi=0$ and $\phi=1$, which define the values
$\phi_0$ and $\phi_{\infty}$ which we considered before in \eqref{bcond}. In this case, the first order equations \eqref{fophi} admit the solution
\be\label{solphi}
\phi(r) = 
\begin{cases}
\tanh^3\left(\frac{1}{3r} - \frac{1}{3r_0}\right);\,\,\,& r \leq r_0\\
0, \,\,\, & r>r_0,
\end{cases} \\
\ee
where $r_0$ is an integration constant, used to control the size of the solution. It connects the aforementioned minima in the compact space $r\in[0,r_0]$. Notice that in the limit $r_0\to\infty$, the solution looses its compact profile and becomes $\phi_\infty(r) = \tanh^3(1/3r)$.

\begin{figure}[t!]
\centering
\includegraphics[width=4.2cm]{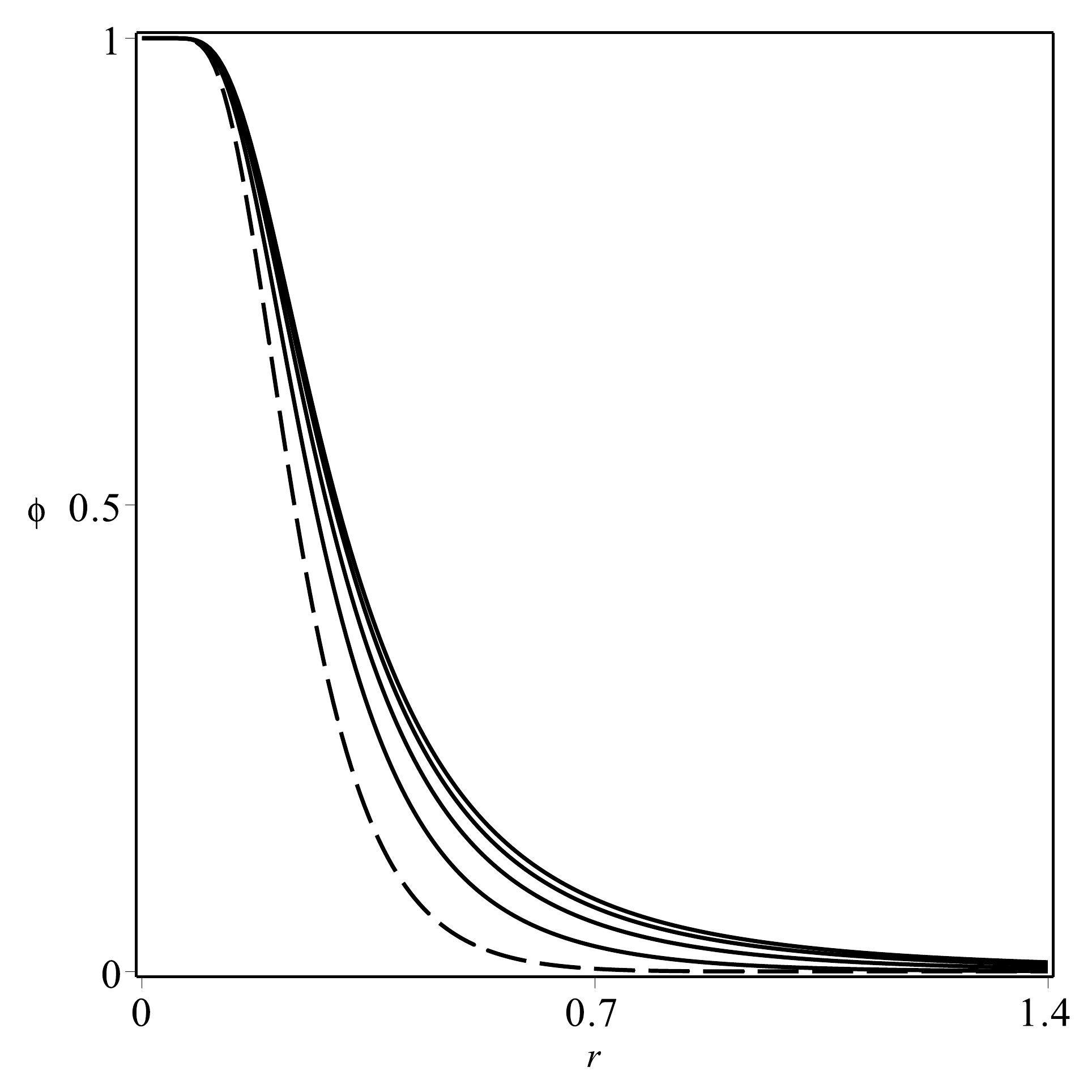}
\includegraphics[width=4.2cm]{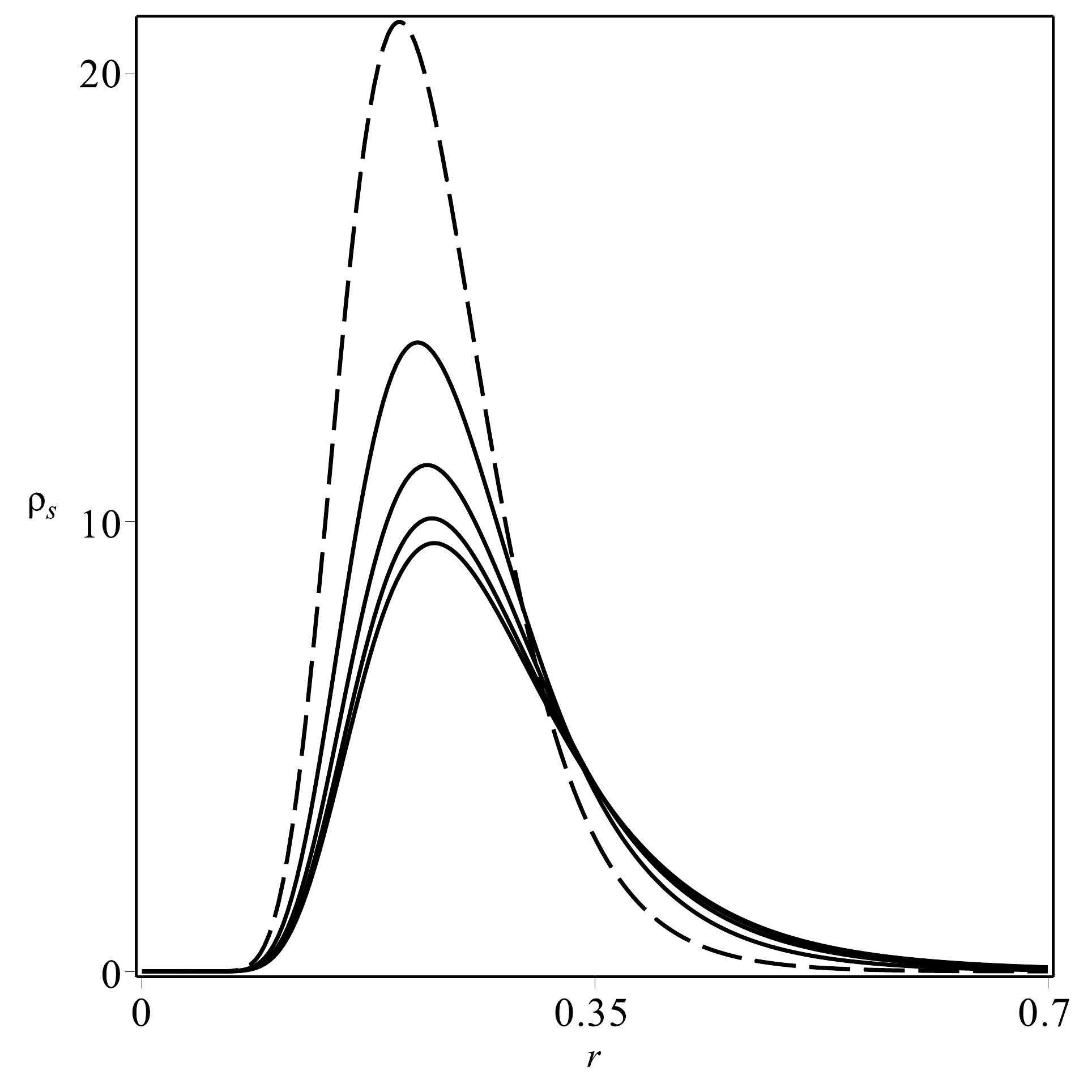}
\caption{The solution in Eq.~\eqref{solphi} (left) and its energy density \eqref{rhophi} (right) for $r_0=1$ (dashed line) and for $r_0=2,4,8,$ and $16$.}
\label{fig4}
\end{figure}

The energy density can be calculated from Eq.~\eqref{rhos}; the result is 
\be\label{rhophi}
\rho_s(r) = 
\begin{cases}
\frac{1}{r^4}\tanh^4\!\left(\!\frac{1}{3r}\! -\! \frac{1}{3r_0}\!\right) \sech^4\!\left(\!\frac{1}{3r}\! -\! \frac{1}{3r_0}\!\right);& r \leq r_0\;\;\;\\
0; & r>r_0,
\end{cases} \\
\ee
which also belongs to the compact space $[0,r_0]$ for $r_0$ finite.  It is possible to check that, by integrating the above equation one gets the energy $E_s=24\pi/35$, which is also the value obtained via the Bogomol'nyi bound in Eq.~\eqref{ebogo}. In Fig.~\ref{fig4}, we depict the solution \eqref{solphi} and its energy density \eqref{rhophi} for some values of $r_0$, and note that the solution presents a plateau near the origin. This makes the energy density of the source field to exhibit a valley in this region. As $r_0$ increases, both the solution and the energy density get wider, with the amplitude of the latter diminishing; this behavior keeps the energy of the source field fixed.

We now use the solution \eqref{solphi} as a source to generate the medium in which monopoles may appear. We suggest two models with distinct behavior and discuss the presence of compact structures. Since the source field solution \eqref{solphi} is compact, we could think of generating compact monopole structures, regardless of the form of the magnetic permeability. As we will show below, however, this is not always the case. The first example we consider is described by the same permeability used before in Eq.~\eqref{perm}. 
By remembering the form of the source field in Eq.~\eqref{solphi}, we see that $P(\phi)=2$ for $r=0$ and $P(\phi)=1$ for $r\geq r_0$. This is not enough to make the monopole compact in this case. In particular, it is straightforward to see that, for $r\geq r_0$, the behavior is as in the standard case, with $P(\phi)=1$, whose solutions are given by $H_{std}(r) = \coth(r)-1/r$ and $K_{std}(r) = r\,\textrm{csch}(r)$, as shown in Ref.~\cite{PS}.

\begin{figure}[t!]
\centering
\includegraphics[width=4.2cm]{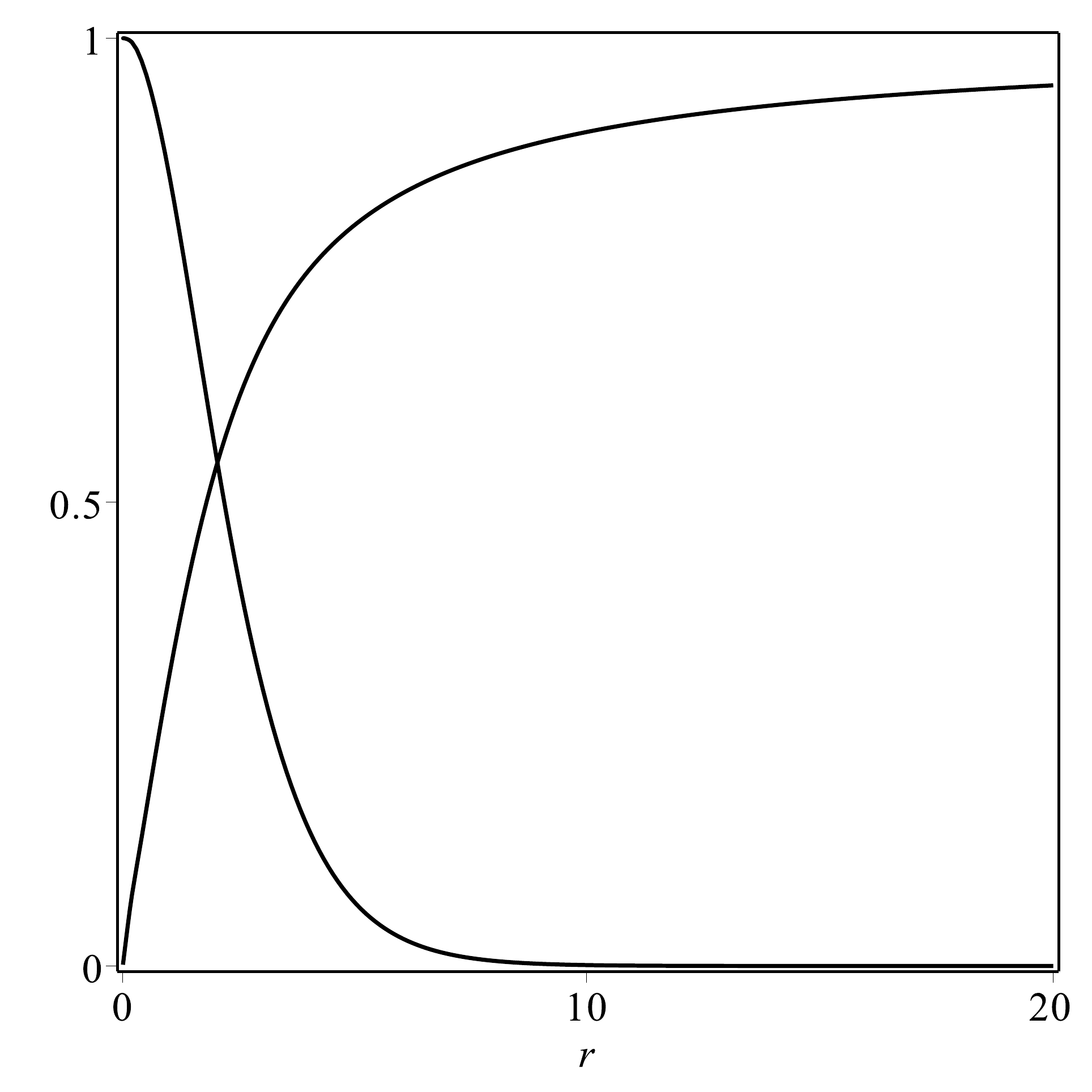}
\includegraphics[width=4.2cm]{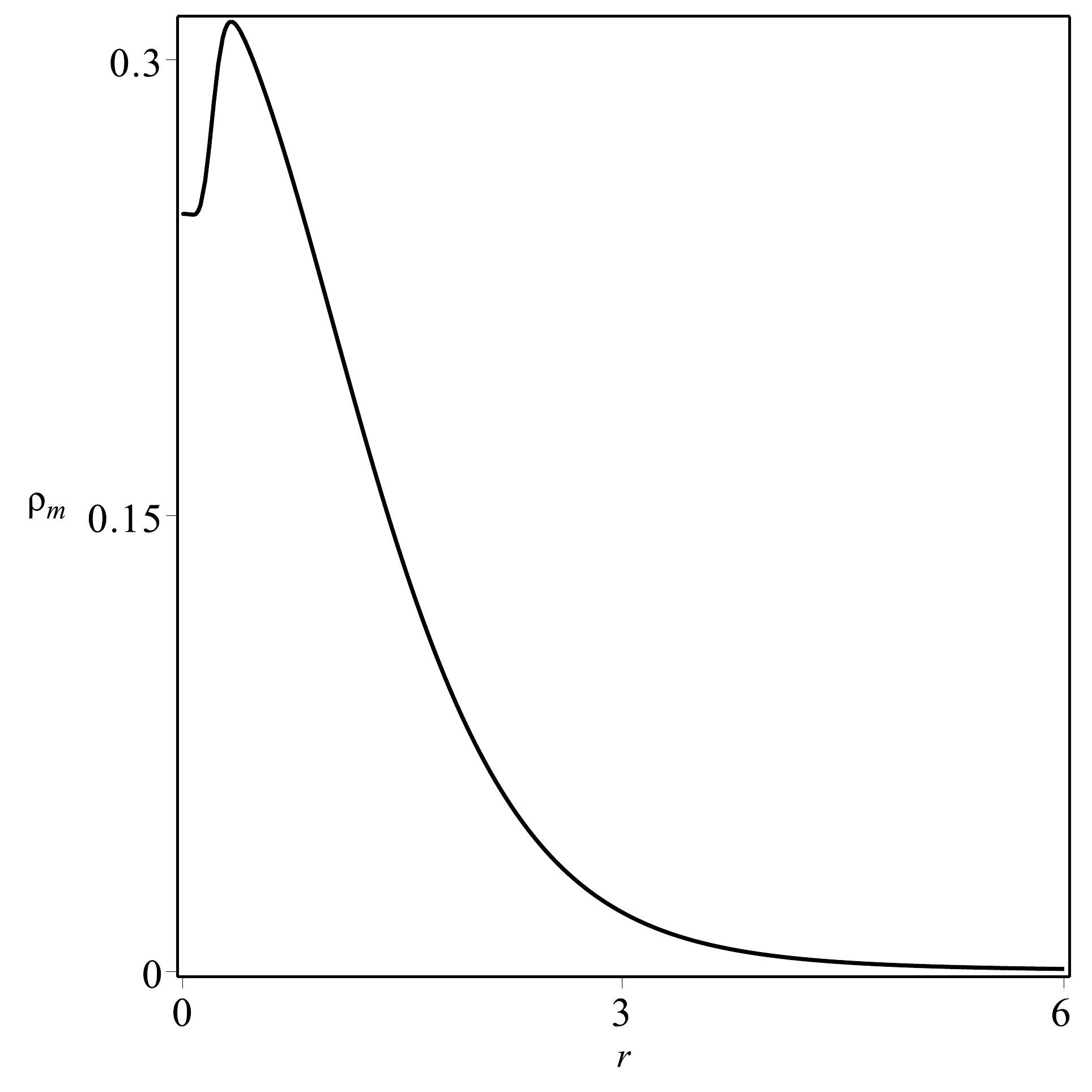}
\caption{The solutions $H(r)$ (ascending line) and $K(r)$ (descending line) of Eqs.~\eqref{fom1} (left) with the source field given by \eqref{solphi} and the energy density of the corresponding monopole (right), for $r_0=1$.}
\label{fig5}
\end{figure}

The above choice for the magnetic permeability changes the first-order equations \eqref{fom} to the form given above by Eq.~\eqref{perm}, but now the neutral field is described by Eq.~\eqref{solphi}. The solutions are depicted in Fig.~\ref{fig5}, together with the energy density. We have considered only $r_0=1$, because this parameter does not modify the solutions significantly. The energy density of the monopole starts increasing abruptly near the origin, and then decreases toward zero asymptotically. This is different from the behavior found for the solutions presented before. A numerical integration all over the space shows that the monopole presents energy $E_m=4\pi$, which matches with Eq.~\eqref{ebogo} for $\eta,g=1$.

\begin{figure}[t!]
\centering
\includegraphics[width=4.6cm]{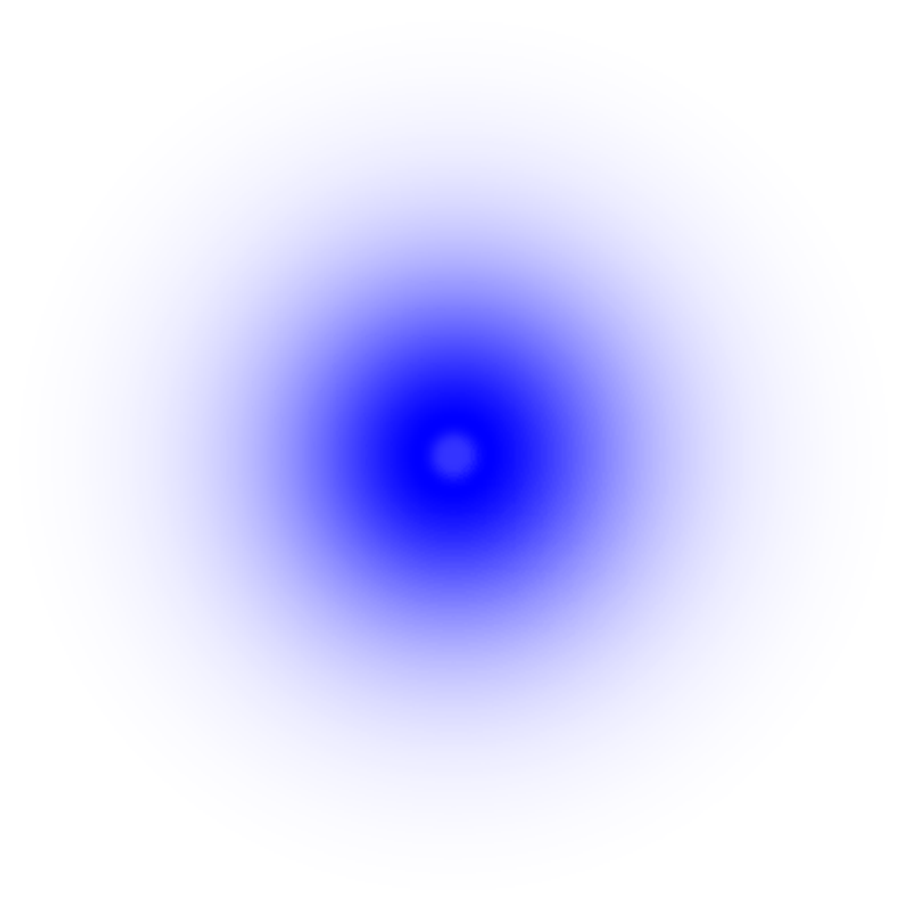}
\caption{The monopole energy density of the solution of Eqs.~\eqref{fom1} with the source field given by Eq.~\eqref{solphi}, for $r_0=1$. We depict a planar section similar to that of Fig.~\ref{fig3}.}
\label{fig6}
\end{figure}

The specific behavior of the source field and the magnetic permeability modifies the internal structure of the monopole, as we can see in Fig.~\ref{fig5}. This motivated us to display in Fig.~\ref{fig6} the energy density of the monopole in a planar section that crosses the center of the structure. We notice a slightly clearer region around the center of the monopole in Fig.~\ref{fig6}, which reflects the fact that the energy density in Fig.~\ref{fig5} starts at a given value and then increases abruptly before decreasing toward zero asymptotically. This identify the presence of internal structure.

Another example arises with the magnetic permeability given by
\be
P(\phi) = \phi^2.
\ee
In this case, since the source field is compact, given by Eq.~\eqref{solphi}, the magnetic permeability vanishes for $r\geq r_0$. As we show below, this gives rise to interesting new features. The first order equations \eqref{fom} become
\bes\label{fom2}
\bal
H^\prime &= \frac{\phi^2(1-K^2)}{r^2},\\
K^\prime &= -\frac{HK}{\phi^2},
\eal
\ees
where $\phi$ is given by Eq.~\eqref{solphi}. Here, since the magnetic permeability vanishes outside the compact space, one may wonder if the compact profile of the source field makes the monopole solutions become compact. Indeed, by investigating their behavior for $r\approx r_0$, we found that $1-H(r)\propto (r_0-r)^7$ and $K(r)\propto \exp(-C/(r_0-r)^5)$, for $C$ real and positive. Therefore, the above equations support solutions that attain their boundary conditions at $r=r_0$, living only in the compact space $r\in[0,r_0]$.

\begin{figure}[t]
\centering
\includegraphics[width=4.2cm]{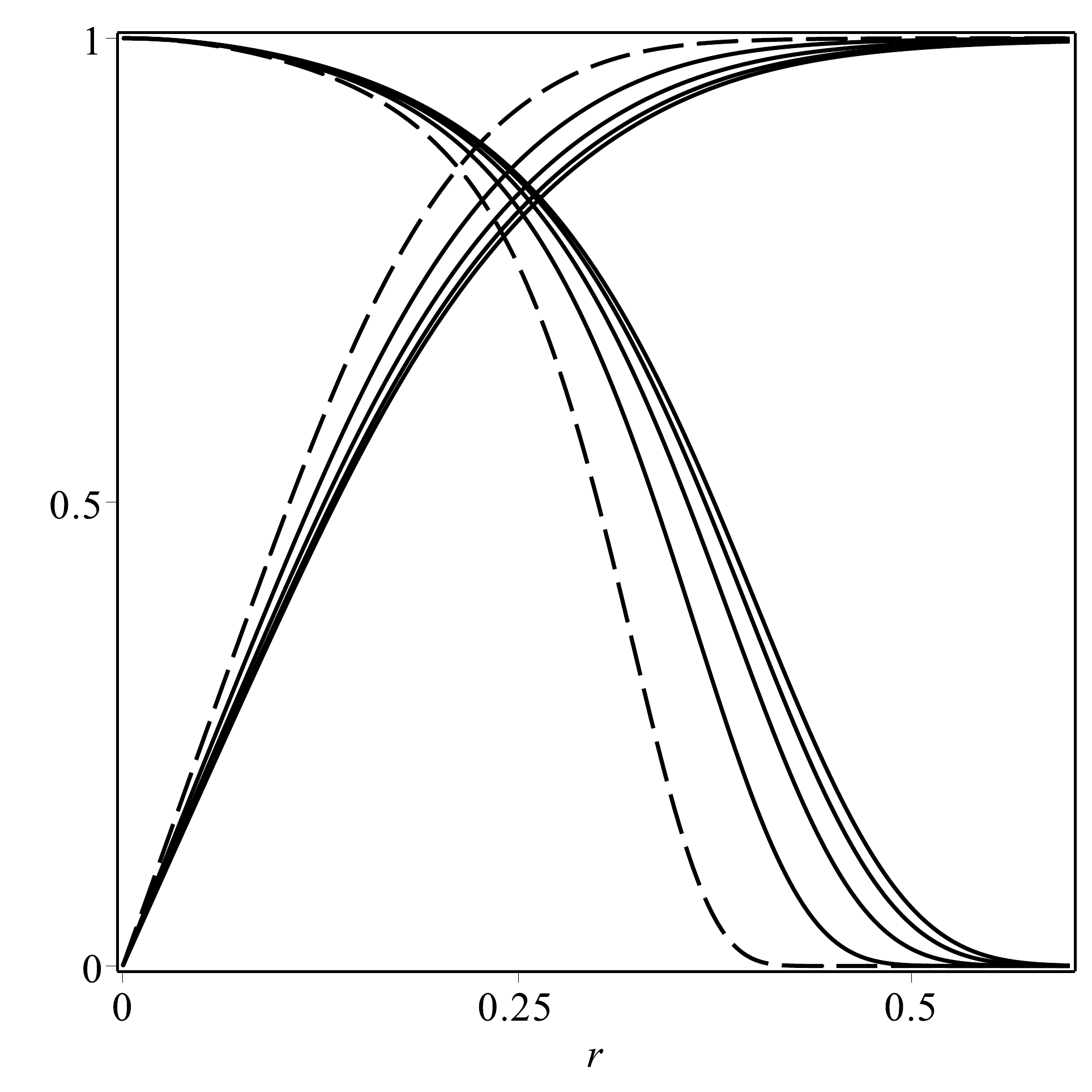}
\includegraphics[width=4.2cm]{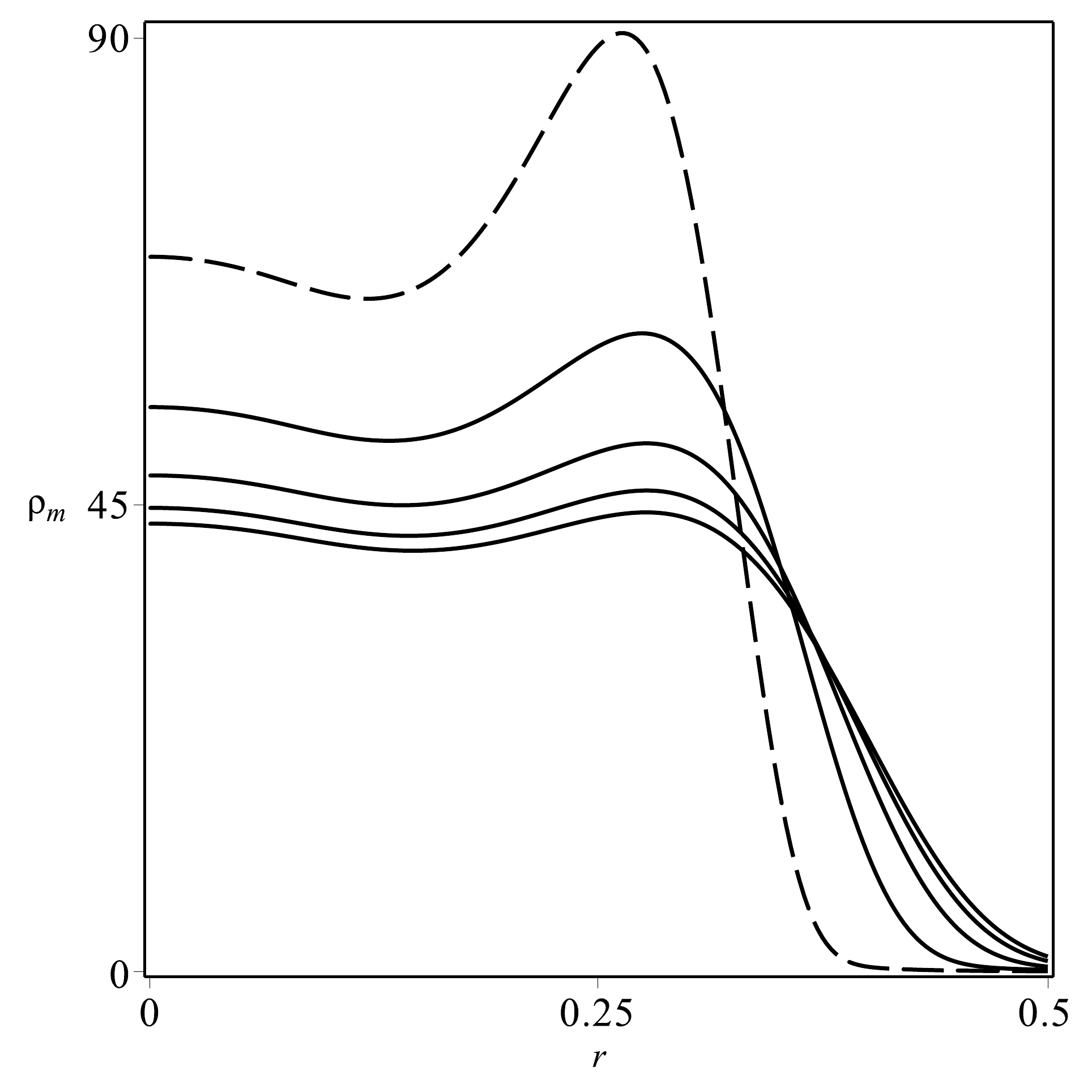}
\caption{The solutions $H(r)$ (ascending lines) and $K(r)$ (descending lines) of Eqs.~\eqref{fom2} (left) with the source field given by \eqref{solphi} and the energy density of the corresponding monopole (right), for $r_0=1$ (dashed) and for $r_0=2,4,8,$ and $16$.}
\label{fig7}
\end{figure}

Unfortunately, we have not been able to find analytical solutions for the above equations \eqref{fom2}. A numerical integration all over the space shows that the monopole presents energy $E_m=4\pi$, as in Eq.~\eqref{ebogo} for $\eta,g=1$. Also, in Fig.~\ref{fig7}, we display the compact solutions and the energy density for some values of $r_0$. The energy density is also compact, vanishing outside the interval $r\leq r_0$. It is also worth mentioning that, as in the source field, the parameter $r_0$ controls the width of the solutions and the energy density, in the sense that the limit $r_0\to\infty$ removes the compact behavior of the system.

\begin{figure}[t!]
\centering
\includegraphics[width=2.8cm]{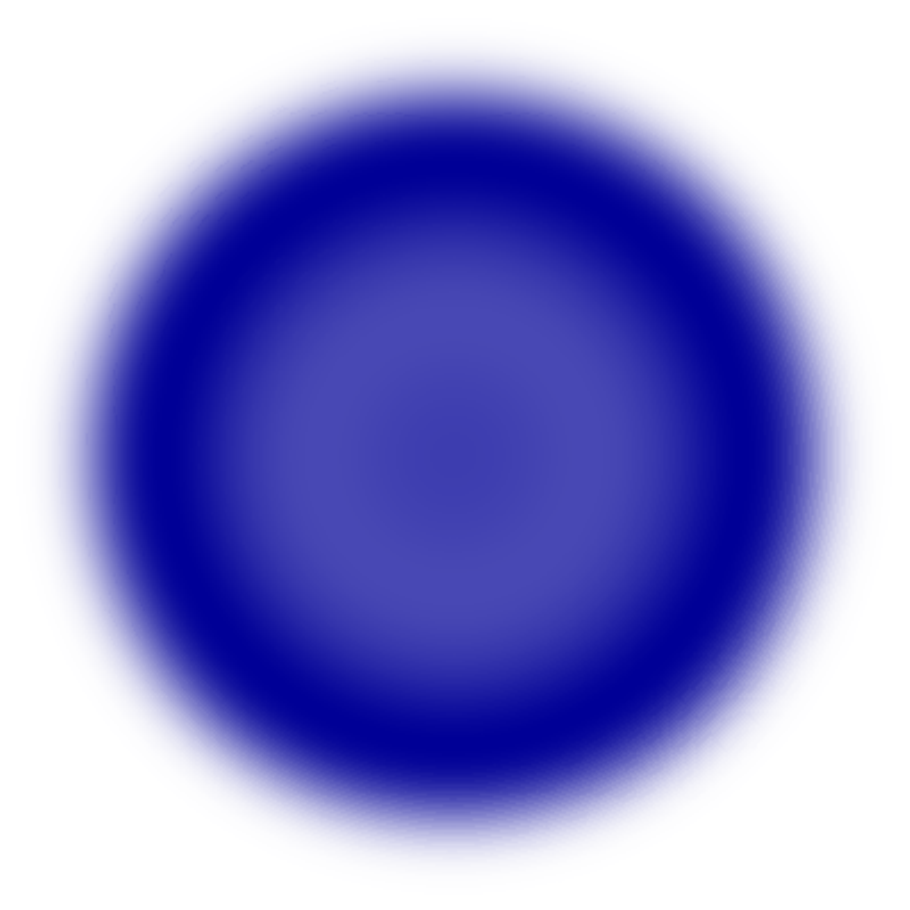}
\caption{The monopole energy density of the solution of Eqs.~\eqref{fom2} with the source field given by Eq.~\eqref{solphi}, for $r_0=1$. We depict a planar section similar to that of Figs.~\ref{fig3} and \ref{fig6}.}
\label{fig8}
\end{figure}

One may notice from the energy density in Fig.~\ref{fig7} that the magnetic monopole also engenders internal structure, which becomes more apparent as $r_0$ decreases. For this reason, in Fig.~\ref{fig8}, we depict a planar section of the energy density of the monopole passing through its center for $r_0=1$. Clearly, it displays a rich internal structure, and the compact profile.

\section{Ending comments}
In this work we studied magnetic monopoles in an enlarged model, in which the $SU(2)$ gauge symmetry is extended to accommodate an extra neutral scalar field that evolves under the $Z_2$ symmetry. In this new model, we also added a generalized magnetic
permeability and a modification in the dynamics of the scalar fields that evolve under the $SU(2)$ symmetry.

As we have shown, the BPS procedure to minimize the energy of the solutions has led us to first order equations that solve the equations of motion. To achieve the first order equations, however, we had to include a specific radial factor into the potential of the neutral field, in a manner similar to that proposed before in \cite{bmmprl} to circumvent the Derrick-Hobart theorem. The mechanism is similar to the case of planar vortices investigated recently in \cite{bmm}, and here it allowed us to find magnetic monopoles with internal structure and compact profile.

We think that the novel results of the work may find applications of current interest both in high energy physics and in condensed matter. In particular, since the BPS procedure involves first order equations one can ask for supersymmetric extensions of the model here considered, to see if supersymmetry works to validate the results of the bosonic model. Another issue concerns the inclusion of fermions, to investigate the presence of fermionic states attached to the monopole. Moreover, we can enlarge the model and add other scalar and gauge fields. These to possibilities are of current interest since they may lead us with electrically charged monopoles, as the magnetic monopoles that appeared in spin ice with an electric dipole \cite{mmed}. Some of these issues are currently under consideration, and we hope to report on them in the near future.

\acknowledgements{We would like to acknowledge the Brazilian agency CNPq for partial financial support. DB thanks support from grant 306614/2014-6, MAM thanks support from grant 140735/2015-1 and RM thanks support from grant 306826/2015-1.}


\end{document}